
\input harvmac.tex

\def\tr{\mathop{\rm tr}\nolimits}
\def\frac#1#2{{\scriptstyle{#1 \over #2}}}

\def\pl#1{Phys.~Lett.~{\bf #1}}
\def\prl#1{Phys.~Rev.~Lett.~{\bf #1}}

\def\s{\phantom{-}}
\tolerance=10000
\hfuzz=5pt
\overfullrule=0pt

\Title{\vbox{\baselineskip12pt\hbox{TPI-MINN-93/9-T}
\hbox{hep-th/9303031}\hbox{March, 1993}}}
{\vbox{\centerline{Is induced QCD really QCD?}
\centerline{The preservation of asymptotic freedom}
\centerline{by matter interactions}}}
\centerline{J.M.~Cline and S.~Paban}
\centerline{Theoretical Physics Institute}
\centerline{University of Minnesota}
\centerline{Minneapolis, Minnesota 55455}
\vskip .3in

The program of induced QCD requires that there exist self-interactions
among the heavy matter fields (an adjoint scalar and a few fermions in the
fundamental representation) which tend to spoil the asymptotic freedom of SU(N)
gauge theory.  We consider general interactions between such matter fields and
show that, on the contrary, to two loops,
they all tend to enhance asymptotic freedom in the
perturbative regime.  This result casts doubt on whether induced QCD is
equivalent to real QCD.
\Date{}

\newsec{Introduction}

Kazakov and Migdal \ref\kaz{V.A.~Kazakov and A.A.~Migdal, Princeton University
preprint PUPT-1322, (1992).}\ have proposed a lattice gauge model of scalar
fields which is solvable in the limit of large $N$, and is supposed to induce
$SU(N)$ gauge theory at energies much smaller than the scalar mass
\ref\mig{A.A.~Migdal, Princeton Univerisity preprint PUPT-1332-REV, (1992);
Ecole Normale Superieure preprint LPTENS-92-22 (1992).}. The idea is to couple
a gauge field with no kinetic term, hence infinite bare gauge coupling
constant, to a heavy scalar field.  When the scalar is integrated out, all
possible local, gauge invariant operators for the gauge vector field will be
induced. Provided that the lowest dimension operator $\tr F^2$ dominates, pure
Yang-Mills theory will have been recovered.  The trick of Kazakov and Migdal is
to reverse the order of integration, since there is no kinetic term for the
gluons, the integral over these fields can be easily performed. The
resulting scalar field theory can be solved in the limit of large $N$.

It remains unproved by them or subsequent workers whether there exists any
choice of parameters in the bare action for which the effective scalar theory
is equivalent to the desired nonabelian gauge theory.  The original and
simplest
version of induced gauge theory was shown to have two difficulties: (1) there
was a spurious $Z_N$ symmetry \ref\kogan{I.I.~Kogan, G.W.~Semenoff and
N.~Weiss,  \prl{69}, 3435 (1992).}\ which caused the Wilson loop operator to
have vanishing expectation value, and (2) the higher dimension operators in the
effective gauge field action (after integrating out the scalar) were not
suppressed relative to $\tr F^2$, resulting in a strongly coupled model similar
to the confining phase of lattice $U(1)$ theory \ref\gross{D.J.~Gross,
\pl{B293}, 181 (1992).}.  Both of these problems can potentially be solved by
adding $N_f$ fermions in the fundamental representation \ref\mixed{A.A.~Migdal,
Ecole Normale Superieure preprint LPTENS-92-23 (1992); Princeton University
preprint PUPT-1343 (1992).}. (An alternative solution to the first problem is
the use of unconventional observables \kogan, \ref\khokh{S.~Khokhlachev, Yu.
{}~Makeenko, ITEP preprint ITEP-YM-5-92 (1992). \semi I.I.~Kogan, A.~Morozov,
G.W.~Semenoff and N.~Weiss, British Columbia University preprint UBCTP 92-26,
ITEP-M6/92 (1992); \semi M.I.~Dobroliubov, I.I.~Kogan, G.W.~Semenoff and
N.~Weiss, British Columbia University preprint UBCTP 92-032; Princeton
University preprint PUPT-1358 (1992).}.) Just one fermion will break the $Z_N$
symmetry, but it would require many (order $N$) to suppress the higher
dimension operators.

Unfortunately, the exact solution of the theory using saddle point methods is
applicable only if $N_f\ll N$.  It was therefore proposed by Migdal \mixed\
to replace large $N_f$ by a large value for the self-coupling $\lambda$ of the
scalar field.  It was hoped that for large enough $\lambda$, the asymptotic
freedom of the effective theory could be spoiled starting at a scale $M$, the
mass of the heavy scalars.  In this way the gauge coupling could run from
infinity at the lattice cutoff down to a small value at $M$, below which the
theory would look like QCD.

We have investigated whether the nongauge couplings of induced QCD can have the
desired effect of spoiling asymptotic freedom and find that, to two loops, the
answer is no. These couplings either leave the beta function of the gauge
coupling $\beta_g$ unchanged, or else make it more negative.   To further
motivate this investigation, in sect.~2 we discuss the possible structure of
the phase diagram and the running of the gauge coupling of induced QCD.
Perturbation theory in the matter couplings can determine the phase boundary in
a small region, which may indicate how the boundary goes at the strong matter
couplings of interest to Migdal's program.  The perturbative contributions to
$\beta_g$ are presented in sect.~3, and we speculate on the meaning of induced
QCD in the final section.

\newsec{Phase diagram of induced QCD}

The operation of integrating out the nonpropagating gauge field of induced QCD,
crucial to solving the theory, is well-defined only using a lattice regulator.
However we will not be interested here in solving the theory, but rather in
discusssing its relation to ordinary QCD.  Since we do not integrate out the
gluons separately, we are free to consider the continuum version of the theory,
using dimensional
regularization.  The Lagrangian density, renormalized at a
running scale $\mu$, is
        \eqn\lagrangian{{\cal L} = -{1\over 4g^2}\tr F^2 + \tr((D_\mu\Phi)^2 -
        M^2\Phi^2) +\sum_j\overline\Psi_j(i\Dsl-M_f)\Psi_j
        -V(\Phi,\Psi),}
with potential
        \eqn\potential{V(\Phi,\Psi) =
        \lambda_1 (\tr(\Phi^2))^2 + \lambda_2 \tr(\Phi^4)
        +\sum_j \overline\Psi_j(y_s+iy_p\gamma_5)\Phi\Psi_j.}
$\Phi$ and the $N_f$ flavors of $\Psi_j$ are taken to be in the adjoint and
fundamental representations, respectively, with masses presumably of the same
order of magnitude.  For simplicity we assumed that the Yukawa couplings are
diagonal.  We might also have included a cubic coupling $\tr(\Phi^3)$ in
\potential, but due to its superrenormalizability it would make no
contribution to the running of $g$. Demanding that \lagrangian\ matches onto
the induced QCD lagrangian at the lattice scale $1/a$, we have the condition on
the gauge coupling that
        \eqn\condition{g(1/a) = \infty.}
On the other hand, $g$ should appear to be asymptotically free at low energies
if induced QCD is QCD. This leads to the picture, Fig.~1, of the hoped-for
behavior of $g$ as a function of $\Lambda$:  it reaches a minimum in descending
from the lattice cutoff to the scale of the heavy particles.  At this point it
starts growing toward the nonperturbative value which it reaches at the
confinement scale.

An obvious way to get this turn-around of $g$ at the intermediate scale $M\sim
M_f$ is to let the number of fermions (or scalars) be large, $N_f \gg N$, for
then the one-loop beta function is given by
        \eqn\betafunc{\beta_g = -{g^3\over 16\pi^2}\left(\frac{11}{3}N -
        \frac{2}{3}N_f - \frac{1}{6}N\right).}
When $N_f$ exceeds the critical value $N_c= 21N/4$, $\beta_g$ becomes positive
and $g$ grows at higher scales.  At scales much less than
$M_f$, the fermions will not
contribute to the running of $g$, so that $\beta_g$ becomes negative and $g$
can also grow in the infrared.   The low energy Lagrangian density has the form
        \eqn\lowenergy{ N_f\left({-1\over 4g^2}\tr F^2 + {c\over M_f^4} \tr F^4
        +\cdots\right),}
where $g,c\sim\ln^p(aM_f)$ are the couplings induced through renormalization.
When $N_f \sim 1$, the higher dimension operators cannot be neglected because
they are inserted into divergent loop diagrams that are cut off at the scale
$M_f$ (where the gluon propagators begin to ``open up'' and reveal their origin
due to the virtual heavy particles).  The higher operators thus all make
equally important contributions, and the theory is very strongly coupled,
resulting in the dynamical generation of a gluon mass, strong screening of the
gauge force, and absence of a continuum limit.  But when $N_f$ is large, the
higher operators are systematically suppressed, as can be seen by the rescaling
$F\to F/N_F^{1/2}$.

Unfortunately, the model can be solved only in the limit $N_f\ll N$.  It was
hoped that the reversal of the the sign of $\beta_g \equiv dg/d({\rm ln}\mu)$
at short distances could instead be accomplished by a suitable choice of matter
couplings, generically denoted by $\lambda$.  For example, the diagram with two
scalar loops in Fig.~2 could in principle make a negative contribution to
$\beta_g \sim g^3\lambda/(4\pi)^4$.  Although this is formally smaller than the
one loop contributions, the number of fermion flavors could be tuned to the
critical value $N_c$ where these vanished, so that the two-loop result would
dominate. Going to slightly smaller values of $N_f$ would make $\beta_g$
negative again, but this could be counteracted by increasing $\lambda$.  Thus
we expect a line of critical points where $\beta_g$ vanishes in the
$\lambda$-$N_f$ plane, which perturbation theory can reliably explore in the
vicinity of $N_f\cong N_c$. Such an exploration is the purpose of this paper.
Suppose a phase boundary exists and extends all the way to $N_f = 0$, as in
Fig.~3a, where the critical value of the coupling $\lambda_c$ is presumably
large.   Then if $\lambda>\lambda_c$, the induced theory may correspond to QCD,
since the gauge coupling runs the right way.  Large values of $\lambda$ pose no
problem for the saddle-point evaluation of the effective scalar theory
resulting from integrating out the gluons. However if $\lambda<\lambda_c$, we
certainly do not expect the induced theory to be the same as QCD.

Although we can say nothing rigorous about the region $N_f\ll N$, since
$\lambda$ will be nonpertubative there, we can make a plausibility argument.
If the phase boundary points toward smaller values of $N_f$ as $\lambda$
increases, it is natural to expect this boundary to exist near $N_f=0$, which
is the hoped-for behavior.  If on the other hand it actually points toward {\it
larger} values of $N_f$, then we would be surprised if any critical value of
$\lambda$ exists near $N_f=0$.  The boundary would have to turn back on itself,
as in Fig.~3b.

\newsec{Two-loop beta function}

We shall now present our results for the contributions to the gauge coupling
beta function coming from the scalars self-couplings and the Yukawa couplings.
These all begin at the two loop order.  First consider the effect of the
quartic self-couplings of  the adjoint scalar field.  At first one might expect
the diagram in Fig.~2 to  make a contribution, but it is easily seen that this
vanishes because the  quartic couplings are symmetric in color indices, whereas
the coupling to  gluons is antisymmetric.  The only possibility then is the
Yukawa coupling  $y = y_s+iy_p\gamma_5$  of the scalar to the fermions. We have
done the computation in the background field gauge, where it is only necessary
to compute the gluon wave function renormalization $Z_3$,

\eqn\renorm{g_{\rm bare} = {Z_3}^{-1/2} g_{\rm ren}.}

There are eight diagrams
which contribute $O(g^3 y^2)$ to $\beta_g$, shown in Fig.~4.  The first four
are one-loop digrams containing counterterms generated by the one-loop
divergences. The last four are two-loop diagrams.  The Feynman rules for the
bare interactions and for the counterterms which we calculated are given in
Fig.~5.  We have used  dimensional regularization with $d=4-2\epsilon$ and
minimal subtraction of the divergent parts of the loops.

To facilitate the calculation, we have computed the trace over Lorentz indices
of the vacuum polarization tensor for each diagram $x$:
        \eqn\pol{\eta^{\mu\nu}\Pi_{\mu\nu}^{ab} = \sum_{x=a,b,\dots h}
        (3- 2 \epsilon) C_x p^2 \delta_{ab},}
where $\eta_{\mu\nu}$ is the Minkowski metric with signature $(+---)$,
and our convention for the sign of $\Pi_{\mu\nu}^{ab}$ is that
        \eqn\polsign{\sum{\rm Figs.\ 4(a-h)\ } = i\Pi_{\mu\nu}^{ab} =
        i\sum_x C_x (\eta_{\mu\nu}p^2 - p_\mu p_\nu)\delta_{ab}.}
The results for the coefficients $C_x$ are given in Table 1, where we show the
contributions to the single and double poles in $\epsilon$.  These results
satisfy a number of consistency checks.  Although for simplicity we assume a
common mass for the fermions and the scalar in Table I, we have checked that
the mass dependence disappears from the sum of diagrams for the divergent
parts.  Furthermore the total contribution to the double pole in $\epsilon$
vanishes.  This is linked to the absence of mass-dependence in the beta
function, which should be the case since we are using a mass-independent
subtraction procedure.  Finally, the the total contributions from the
scalar and pseudoscalar couplings have the same strength, even though they
differ from diagram to diagram.  This must be the case for any
mass-independent quantity, since in the absence of a mass term one can
transform between scalar and pseudoscalar couplings.

\midinsert\noindent {\ninerm Table 1: Contributions to the vacuum polarization
tensor, to be multiplied by the factor $g^2 (y_s^2+y_p^2) N N_f/(3(4\pi)^4)$,
where $\theta_s^2 = y_s^2/y^2$, $\theta_p^2 = y_p^2/y^2$ and $y^2 =
y_s^2+y_p^2$, parametrizing the contributions of the scalar and pseudoscalar
couplings, respectively.  $X$ is
$(\Gamma'(1)/\Gamma(1)-\ln(M^2/(16\pi^2\mu^2))$, which along with the $\ln(2)$
contributions of the last column is an artifact of dimensional regularization
that disappears from the total result.}
$$\vbox{\tabskip=0pt \offinterlineskip
\halign to \hsize{\strut#& \vrule#\tabskip 1em plus 2em minus .5em&
\hfil#\hfil &\vrule#& \hfil#\hfil &\vrule#& \hfil#\hfil &\vrule#&
\hfil#\hfil &\vrule#& \hfil#\hfil &\vrule#& \hfil#\hfil &\vrule#\tabskip=0pt\cr
\noalign{\hrule}
&& Diagram && $1/\epsilon^2$ && $\theta_s^2/\epsilon$ &&
$\theta_p^2/\epsilon$
&& $X/\epsilon$ && $\ln(2)/\epsilon$ &\cr
\noalign{\hrule}
&& Fig.~4a && $\s 1$  && 0        && $\s 0$ && $\s 1$ && $-1  $ &\cr
&& Fig.~4b && $-1  $  && 3        && $- 1$ && $- 1$ && $\s 1$ &\cr
&& Fig.~4c && $\s 1$  && 0        && $\s 0$ && $\s 1$ && $\s 0$ &\cr
&& Fig.~4d && $-1  $  && $-5/2$  && $-1/2$ && $-1  $ && $\s 0$ &\cr
\noalign{\hrule}
&& Fig.~4e && $1/2$ && $19/6$   && $7/6$   && $\s 1$  && $-1$   &\cr
&& Fig.~4f && $1/2$ && $-25/12$ && $23/12$ && $\s 1$  && $-1$   &\cr
&& Fig.~4g && $-1$  && $-5/6$   && $-5/6$  && $-2  $  && $\s 2$ &\cr
&& Fig.~4h && $O(1/N)$ && $O(1/N)$ && $O(1/N)$ && $O(1/N)$ && $O(1/N)$ &\cr
\noalign{\hrule}
&& Total   &&  0    && $3/4$    && $3/4$   &&  0   &&  0   &\cr
\noalign{\hrule}
}}$$
\endinsert
The contribution to the gauge coupling beta function from the Yukawa
interactions can be computed from our results for the vacuum polarization.
Denoting the bare coupling by
        \eqn\bare{g_0 = \mu^{\epsilon}
g\left(1+c{g^2 y^2\over\epsilon}\right)}
and demanding it be independent of the subtraction point, $(\mu d/d\mu)g_0 =0$,
we find that $\beta_g = 4cg^3y^2$, where use has been made of the fact that
$\beta_y = \epsilon y +O(\epsilon^0)$, due to the dimensionality of the bare
$y$ in $4-2\epsilon$ dimensions.  From Table I, and $\polsign$ we see that
$c = -g^2 y^2 N N_f /(8(4\pi)^4))$.  Let us keep careful track of the signs.
There are three factors of $-1$, one from the sign in the kinetic term for the
gluons, eq.~\lagrangian, one from the fact that the counterterm must have the
opposite sign to the divergent loop calculation, and one from the fact that
$1/Z_3$ rather than $Z_3$ appears in eq.~\renorm.
So the shift in $\beta_g$ is
        \eqn\final{\Delta\beta_g = -{g^3y^2NN_f\over 2(4\pi)^4}.}
The sign of \final\ is our main result.  It indicates that the curvature of the
phase boundary near $N_f=N_c$ and zero coupling is as in Fig.~3b, not 3a.

\newsec{Conclusions}

We have shown that perturbatively, the effect of scalar self-interactions and
Yukawa couplings is to leave unaltered or reinforce the asymptotic freedom of
the gluons.  If this is also true for large values of the couplings, then the
gauge coupling of induced QCD not only starts out at infinity at the lattice
cutoff, but remains there at longer distance scales.  Then the gauge degrees of
freedom would remain frozen, and induced QCD would really just be a theory of
scalar fields.  It has been argued that this theory has a continuum limit,
which would make it interesting in its own right, as the only other such
example aside from nonabelian gauge theory.  We would have no reason to believe
it to be equivalent to the pure gauge theory.

A possible loophole may be provided by the three-loop contribution to $\beta_g$
of order $g^3\lambda^2$.  It would be the leading effect of the scalar quartic
coupling since, as explained above, the two-loop contribution vanishes.

It may also be possible to recover QCD from induced QCD by enlarging the class
of matter field interactions.  We restricted ourselves to the renormalizable
ones, but since the matter fields of induced QCD are not supposed to represent
fundamental, physical particles, it may be sensible to consider higher
dimension operators, such as
        \eqn\hidim{{\lambda\over M^2}\left(\partial_\mu\tr(\Phi^2)\right)^2.}
This interaction should provide a contribution of order $g^3\lambda$ at two
loops.  The sign could be adjusted by choosing the sign of $\lambda$.  Such a
derivative coupling, if important, would change the analysis of the effective
scalar field theory drastically, since the saddle point would no longer be a
homogeneous field configuration.

Finally, it may be that perturbation theory is not a good indicator of the beta
function at strong couplings, so that induced QCD nevertheless is QCD.  If this
is true however, one should be able to see the phase transition in the
effective scalar field theory with the gluons integrated out, when the
couplings $\lambda$ or $y$ fall below some critical values.  This could
constitute a better criterion for whether induced QCD is QCD than the
perturbative one we have investigated.

We would like to thank A.A.~Migdal for suggesting this problem, as well as many
of the important ideas.

\vfill\eject
\nfig\figone{Hoped-for behavior of the running gauge coupling in induced QCD,
where $a$ is the lattice cutoff and $M$ the heavy scalar mass.}
\nfig\figtwo{Two-loop diagram with scalar fields that could affect the running
of the gauge coupling.}
\nfig\figthree{Possible behaviors of the phase boundary between the theory
which should correspond to QCD at large distances (labeled
``not A.F.,'' since the gauge coupling must be trivial in the region between
the lattice cutoff and the heavy scalar mass) and the strongly coupled phase
which one wishes to avoid (labeled ``A.F.'').  The phase boundary is the line
on which the beta function $\beta_g$ vanishes.}
\nfig\figfour{Feynman diagrams contributing to $Z_3$ at order $g^2y^2$.
Double lines are scalars, single lines are fermions, and heavy dots are
counterterm interactions.}
\nfig\figfive{Feynman rules used to calculate diagrams in \figfour.  The
constants are given by $K_{2,\psi} = -N/(64\pi^2\epsilon)$,
 $K_{1,\psi} = gy^2N/(64\pi^2\epsilon)$,
 $K_{1,\phi} = -g/(16\pi^2\epsilon)$, and
 $K_{2,\phi} = -1/(16\pi^2\epsilon)$, where $y^2 = y_s^2+y_p^2$ and $\delta y^2
=y_s^2-y_p^2$.}
\listrefs
\listfigs
\end